\documentclass[traditabstract]{aa} 
\usepackage{graphicx}
\usepackage{txfonts}
\usepackage{natbib}
\bibpunct{(}{)}{;}{a}{}{,} 
\begin{document}
   \title{
The evolving SFR - M$_{\star}$ relation and sSFR since z$\simeq$5 from the VUDS spectroscopic survey
	  \thanks{Based on data obtained with the European 
	  Southern Observatory Very Large Telescope, Paranal, Chile, under Large
	  Program 185.A--0791. }
}

\author{L. A. M.~Tasca\inst{1}
	  \and O.~Le F\`evre\inst{1}
          \and N.P.~Hathi\inst{1}
	  \and D. Schaerer\inst{2,3}
          \and O.~Ilbert\inst{1}
	  \and G.~Zamorani \inst{4}
	  \and B.~C.~Lemaux \inst{1}
\and P.~Cassata\inst{1,5}
\and B.~Garilli\inst{6}
\and V.~Le Brun\inst{1}
\and D.~Maccagni\inst{6}
\and L.~Pentericci\inst{7}
\and R.~Thomas\inst{1}
\and E.~Vanzella\inst{4}
\and E.~Zucca\inst{4}
\and R.~Amorin\inst{7}
\and S.~Bardelli\inst{4}
\and L.~P.~Cassar\`a\inst{6}
\and M.~Castellano\inst{7}
\and A.~Cimatti\inst{8}
\and O.~Cucciati\inst{8,4}
\and A.~Durkalec\inst{1}
\and A.~Fontana\inst{7}
\and M.~Giavalisco\inst{9}
\and A.~Grazian\inst{7}
\and S.~Paltani\inst{10}
\and B.~Ribeiro\inst{1}
\and M.~Scodeggio\inst{6}
\and V.~Sommariva\inst{8,7}
\and M.~Talia\inst{8}
\and L.~Tresse\inst{1}
\and D.~Vergani\inst{11,4}
\and P.~Capak\inst{12}
\and S.~Charlot\inst{13}
\and T.~Contini\inst{14}
\and J.G. Cuby\inst{1}
\and S.~de la Torre\inst{1}
\and J.~Dunlop\inst{15}
\and S.~Fotopoulou\inst{10}
\and A.~Koekemoer\inst{16}
\and C.~L\'opez-Sanjuan\inst{17}
\and Y.~Mellier\inst{13}
\and J.~Pforr\inst{1}
\and M.~Salvato\inst{18}
\and N.~Scoville\inst{19}
\and Y.~Taniguchi\inst{20}
\and P.W. Wang\inst{1}
}

\institute{Aix Marseille Universit\'e, CNRS, LAM (Laboratoire d'Astrophysique de Marseille) UMR 7326, 13388, Marseille, France
\and
Geneva Observatory, University of Geneva, ch. des Maillettes 51, CH-1290 Versoix, Switzerland
\and
Institut de Recherche en Astrophysique et Plan\'etologie - IRAP, CNRS, Universit\'e de Toulouse, UPS-OMP, 14, avenue E. Belin, F31400
Toulouse, France
\and
University of Bologna, Department of Physics and Astronomy (DIFA), V.le Berti Pichat, 6/2 - 40127, Bologna
\and
Instituto de Fisica y Astronom\'ia, Facultad de Ciencias, Universidad de Valpara\'iso, Gran Breta$\rm{\tilde{n}}$a 1111, Playa Ancha, Valpara\'iso Chile
\and
INAF--IASF Milano, via Bassini 15, I--20133, Milano, Italy
\and
INAF--Osservatorio Astronomico di Roma, via di Frascati 33, I-00040, Monte Porzio Catone, Italy
\and
University of Bologna, Department of Physics and Astronomy (DIFA), V.le Berti Pichat, 6/2 - 40127, Bologna
\and
Astronomy Department, University of Massachusetts, Amherst, MA 01003, USA
\and
Department of Astronomy, University of Geneva
ch. d'\'Ecogia 16, CH-1290 Versoix, Switzerland
\and
INAF--IASF Bologna, via Gobetti 101, I--40129,  Bologna, Italy
\and
Department of Astronomy, California Institute of Technology, 1200 E. California Blvd., MC 249-17, Pasadena, CA 91125, USA
\and
Institut d'Astrophysique de Paris, UMR7095 CNRS,
Universit\'e Pierre et Marie Curie, 98 bis Boulevard Arago, 75014
Paris, France
\and
Institut de Recherche en Astrophysique et Plan\'etologie - IRAP, CNRS, Universit\'e de Toulouse, UPS-OMP, 14, avenue E. Belin, F31400
Toulouse, France
\and
SUPA, Institute for Astronomy, University of Edinburgh, Royal Observatory, Edinburgh, EH9 3HJ, United Kingdom
\and
Space Telescope Science Institute, 3700 San Martin Drive, Baltimore, MD 21218, USA
\and
Centro de Estudios de F\'isica del Cosmos de Arag\'on, Teruel, Spain
\and
Max-Planck-Institut f\"ur Extraterrestrische Physik, Postfach 1312, D-85741, Garching bei M\"unchen, Germany
\and
Department of Astronomy, California Institute of Technology, 1200 E. California Blvd., MC 249--17, Pasadena, CA 91125, USA
\and
Research Center for Space and Cosmic Evolution, Ehime University, Bunkyo-cho 2-5, Matsuyama 790-8577, Japan \\ \\
             \email{lidia.tasca@lam.fr}
}

   \date{Received November 2014; accepted ...} 

%
 
  \abstract{
We study the evolution of the star formation rate (SFR) -- stellar mass (M$_{\star}$) relation and specific star formation rate (sSFR) of star forming galaxies (SFGs) since a redshift $z\simeq5.5$ using 2435 (4531) galaxies with highly reliable (reliable) spectroscopic redshifts in the VIMOS Ultra--Deep Survey (VUDS). 
It is the first time that these relations can be followed over such a large redshift range from a single homogeneously selected sample of galaxies with spectroscopic redshifts.
The $log(SFR) - log(M_{\star})$ relation for SFGs remains roughly linear all the way up to $z=5$ but the SFR steadily increases at fixed mass with increasing redshift.
We find that for stellar masses M$_{\star} \geq 3.2 \times 10^{9}$ M$_{\sun}$  the SFR increases by a factor $\sim13$ between $z=0.4$ and $z=2.3$. 
We extend this relation up to $z=5$, finding an additional increase in SFR by a factor $1.7$ from $z=2.3$ to $z=4.8$ for masses M$_{\star} \geq 10^{10}$ M$_{\sun}$.  
We observe a turn--off in the SFR--M$_{\star}$ relation at the highest mass end up to a redshift z$\sim3.5$.
We interpret this turn--off as the  signature of a strong on--going quenching mechanism and rapid mass growth. 
The sSFR increases strongly up to $z\sim2$ but it grows much less rapidly in $2<z<5$.
We find that the shape of the sSFR evolution is not well reproduced by cold gas accretion--driven models or the latest hydrodynamical models. 
Below z$\sim2$ these models have a flatter evolution $(1+z)^{\Phi}$ with $\Phi=2-2.25$ compared to the data which evolves more rapidly with $\Phi=2.8\pm0.2$. Above z$\sim2$, the reverse is happening with the data evolving more slowly with $\Phi=1.2\pm0.1$. 
The observed sSFR evolution over a large redshift range $0<z<5$ and our finding of a non linear main sequence at high mass both indicate that the evolution of SFR and M$_{\star}$ is not solely driven by gas accretion. 
The  results presented in this paper  emphasize the need to invoke a more complex mix of physical processes including major and minor merging to further understand the co--evolution of the star formation rate and stellar mass growth.
}


\authorrunning{Lidia A. M. Tasca et al.}
\titlerunning{Evolution of the SFR--M$_{\star}$ relation and sSFR up to z$\simeq5$ from VUDS}

   \maketitle


\section{Introduction}

Star formation is a process fundamental to galaxy evolution.
Stars are forming from gas made available from accretion processes or which is recycled by exploding stars in their evolution cycles, different stellar populations with different histories are mixed in mergers, and they follow the dynamical evolution of galactic components forming bulges and discs ultimately leading to the spiral and elliptical galaxies observed today. 

It is now well established that the star formation rate (SFR) history went through several phases \citep{Madau:14}. The SFR apparently rose after the first galaxies formed \citep{Bouwens:14}, then reached through a maximum plateau or peak at redshifts z$\sim1.5-2.5$ followed by a sharp decline to the present \citep[e.g.][]{Bouwens:14,Cucciati:12}. 
A main sequence in the star formation rate (SFR) vs. stellar mass (M$_{\star}$)  plane has been identified for star-forming galaxies, and is strongly evolving with redshift \citep{Elbaz:07,Whitaker:12,Fumagalli:13}. 
By redshift $1$ the main sequence is $\sim$7 times higher in star formation rate \citep{Elbaz:07} than the local relation \citep{Brinchmann:04}, and this becomes $\sim$20 times higher by redshift 2 \citep{Daddi:07}. The scatter in this relation can possibly provide interesting constraints on the star formation history \citep{Salmon:14}. 
While a linear main sequence seems to well represent the observations at intermediate masses, claims for departure from a linear relation have been made for the massive end of the distribution \citep{Karim:11, Whitaker:12,Whitaker:14,Schreiber:14} which may indicate faster mass growth than expected from gas accretion alone and/or star formation quenching. 

The redshift evolution of the specific star formation rate $sSFR=SFR/M_{\star}$, M$_{\star}$ being the stellar mass, is the matter of considerable debate.
While it seems agreed that the mean (or median) sSFR $\langle$sSFR$\rangle$ at a given mass is steadily rising back to z$\sim2$ \citep[e.g.][]{Whitaker:12,Fumagalli:13}, the issue is far from being settled at $z>2$. 
Over the past few years, several apparently conflicting measurements either show no evolution with redshift of  $\langle$sSFR$\rangle$ \citep{Stark:09},  a significant rise  \citep{Stark:13,deBarros:14,Salmon:14}, or even a decrease \citep{Bouwens:12}.
The $\langle$sSFR$\rangle$ was originally reported to evolve weakly over $4 < z < 6$ by \citet{Stark:09}, but the same authors then reported from improved SED fitting that the $\langle$sSFR$\rangle$ evolves more rapidly at $z > 4$ than previously thought \citep{Stark:13}.
Their new results support up to a $5\times$ increase in $\langle$sSFR$\rangle$ between $z \sim 2$ and 7, and they claim that such a trend is much closer to expectations from cold gas accretion models \citep[e.g.][]{Dekel2009}. 
The latest study  from the CANDELS survey seems to agree with this picture \citep{Salmon:14}.
Given the different selection functions of samples used in these calculations combined with the uncertainties in deriving M$_{\star}$ and SFR for faint galaxy samples, this situation needs to be clarified with a better understanding from both the observational and simulations side. 

On the observational side, efforts have been made to understand the limitations in computing M$_{\star}$ and SFR.
The derivation of these physical parameters may be performed with a limited number of methods.
At lower redshifts $z <\sim 2$ the SFR is often derived from the rest UV emission which needs to be corrected from dust extinction, or from the far IR luminosity assuming dust is heated by forming stars, as measured with the Herschel space observatory in recent years. 
Other important SFR indicators include spectral analysis for emission lines like [OII]3727, H$\beta$ or H$\alpha$.
The $H\alpha$ line is generally assumed to provide the most direct estimate of the SFR, now measured to $z\sim2$, but at $z>1$ this is done for relatively small samples of bright galaxies \citep[e.g.][]{Silverman:14}.
At higher redshifts $z>2$ the sensitivity of Herschel has enabled to measure $L_{IR}$ on samples of a few hundred galaxies \citep[e.g.][]{Lemaux:13,Rodighiero:14}.
Measuring M$_{\star}$ and SFR is most often done from the same process along with photometric redshift determination, as $z_{phot}$, M$_{\star}$ and SFR may all rely on the same template fitting of the observed photometry distribution over a broad wavelength range. 
There are several difficulties linked to this process, including the impact on M$_{\star}$ and SFR of using different SFH, initial mass functions (IMF) or dust extinction laws. 
These physical parameters depend in particular on the assumed SF histories and age priors as will be discussed in a forthcoming paper \citep{Cassara:14}.
With this in mind, it is generally assumed that different methods and assumptions lead to typical uncertainties on M$_{\star}$ of $\sim0.2-0.3$dex \citep{Madau:14}.
One of the important limitations in measuring M$_{\star}$ and SFR from SED fitting was identified a number of years ago see \citep[see e.g.][]{Ilbert:09}: the photometric magnitudes in some observed bands may be including flux from nebular emission lines in addition to stellar continuum, while reference templates may not include emission lines at all, requiring a specific correction when the SED fitting is performed. 
Emission lines can increase the observed flux in a particular band by up to 1 magnitude or so, particularly in the K--band and Spitzer--IRAC bands for $z>2$ most important to derive M$_{\star}$ from weighted older stellar populations, and specific protocols are implemented to address this problem \citep[e.g.][]{Ilbert:09,deBarros:14}.
Correcting for emission lines contamination lowers M$_{\star}$, and hence increases the sSFR, therefore making significant differences in the measurement of the star formation main sequence at different redshifts.  
In \citet{deBarros:14}, the effect of nebular lines, once corrected, leads to a steeper evolution of the $\langle$sSFR$\rangle$ at $z>3$. 
Considering these limitations and using 25 different studies in the recent literature \citet{Speagle:14} claim that there is ``a remarkable consensus among MS observations'' with a 0.1 dex 1$\sigma$ inter--publication scatter.
One should however remain cautious that the similar methods used in these studies might lead to similar results as they are affected by similar limitations and uncertainties.

The predicted evolution of $\langle$sSFR$\rangle$ from simulations has been addressed from a number of studies.
Models with constant star formation from the continuous accretion of gas in cold flows along the cosmic web \citep{Neistein:08,Dekel2009} show a continuously increasing M$_{\star}$ and the $\langle$sSFR$\rangle$ is expected to evolve with redshift following $(1+z)^{2.25}$ \citep{Dekel2009,Dutton:10}. 
Hydrodynamical simulations also predict a continuously increasing $\langle$sSFR$\rangle$ but at levels systematically lower than the cold accretion models \citep{Dave:11,Sparre:14}. 
The constant $\langle$sSFR$\rangle$ as originally reported by \citet{Stark:09} at $z>2$ would be puzzling in the context of these galaxy--formation models \citep{Weinmann:11} as to reproduce such a trend would require non--trivial modifications to models, including a suppressed SFR at $z>4$ in galaxies of all masses, a delayed build up of stellar mass from streamed gas, or enhanced growth of massive galaxies with a faster assembly or more efficient starbursts in mergers. 
\citet{Weinmann:11} conclude that a constant $\langle$sSFR$\rangle$ at high z would make it difficult to form enough massive galaxies at $z \sim 1-3$ in SAM, unless the rate of mass assembly due to mergers and the associated starbursts are pushed to the model limits. 
Finding a rising $\langle$sSFR$\rangle$ with redshift then seems much more acceptable from a model point of view.

The most recent data seem to indicate that the $\langle$sSFR$\rangle$ continues rising beyond redshift z$\sim2$ \citep{Stark:13,deBarros:14,Salmon:14}. 
However large uncertainties remain in the determination of this relation up to high redshifts z$\sim6$, resulting from both the determination of the physical parameters SFR and M$_{\star}$ and from the small fields observed leading to significant cosmic variance. 
Further exploration of these relations from new independent datasets is therefore in order.

In this paper we use 4531 galaxies with spectroscopic redshifts in the VIMOS Ultra Deep Survey (VUDS), the largest spectroscopic survey available at $2<z<6$ \citep{LeFevre:14}, to investigate the evolution of the SFR--M$_{\star}$ relation and of the mean sSFR over this redshift range. 
The VUDS survey covers 1 square degree in 3 different fields, minimizing cosmic variance effects. We use the Le Phare code for SED fitting, including emission line treatment as described in \citet{Ilbert:09}.  We describe the VUDS spectroscopic data and associated photometric data used in the SED fitting in Section \ref{obs}. The methodology to measure M$_{\star}$ and SFR from SED fitting is described in Section \ref{sed}. We present the evolution of the SFR--M$_{\star}$ relation from redshift z=0.5 to z=5 in Section \ref{sfr_mass}.
The evolution of the $\langle$sSFR$\rangle$ is discussed in Section \ref{ssfr_z}. We conclude in Section \ref{discuss}.

We use a cosmology with $H_0=100h~km~s^{-1}~Mpc^{-1}$, {$h=0.7$}, $\Omega_{0,\Lambda}=0.7$ and $\Omega_{0,m}=0.3$. 
All magnitudes are given in the AB system, and we keep the AB notation apparent throughout the paper.


\section{The VUDS spectroscopic sample}
\label{obs}

The VIMOS Ultra Deep Survey (VUDS) is a spectroscopic survey of  $\approx10\,000$ galaxies to study galaxy evolution in the redshift range $2<z<6+$, as described in \citet{LeFevre:14}. 
Galaxies in this redshift range are selected from a combination of photometric redshifts, with the first or second peaks in the $z_{phot}$ probability distribution function satisfying $z_{phot}+1\sigma>2.4$, as well as from colour selection criteria like LBG, combined with a flux limit $22.5 \leq i_{AB} \leq 25$. 
A random purely flux selected sample with $i_{AB}=25$ is added to the spectroscopic multi--slit masks, geometry permitting.
Spectra are obtained with the VIMOS spectrograph on the ESO--VLT \citep{LeFevre:03}, covering a wavelength range $3650 < \lambda < 9350$\AA ~at a resolution $R\simeq230$, with integration times of $\sim$14h.
Redshifts are measured from these spectra in a well controlled process delivering a reliability flag for each measurement \citep{LeFevre:14}. 
This is the largest spectroscopic sample at this depth and in this redshift range available today.

The redshift distribution of the VUDS sample extends from z$\sim2$ to higher than z$\sim6$ (median $z=3$), and a lower redshift sample is also assembled from z$\sim0$ to z$\sim2$ (median $z=0.9$) coming from the random flux--selected sample.
For this study we use a total sample of 4531 galaxies with a reliable spectroscopic redshift measurement over the whole redshift range $0<z<6$ (this represents about two--thirds of the final sample as data processing is in progress for the last third). 

For this study we use the 2435 galaxies in VUDS with the highest spectroscopic redshift reliability, flags 3 and 4. 
This is the core sample used in the main analyses of the SFR--M$_{\star}$ and sSFR(z) relations presented below. 
We also use flag 2 and 9 objects which are $\sim70-75$\% reliable (as measured from repeated observations) to augment the size of the sample in the highest redshift bin $z>4.5$.

The VUDS survey is conducted in 3 fields, COSMOS, ECDFS, and VVDS--02h (also known as CFHTLS--D1). 
Each of these fields has extensive very deep multi--band photometry ranging {\it at minima} from broad band {\it u} to Spitzer--IRAC 4.5$\mu$m band. The COSMOS field has the most extensive photometric set, with 30 bands including standard broad band as well as medium band photometry \citep[see][and references therein]{Ilbert:13}. 
The ECDFS and CFHTLS--D1 have accumulated exceptional deep broad band photometric datasets, as described in \citet{Cardamone:10} and \citet{LeFevre:14} respectively.


\section{Spectral energy distribution fitting: M$_{\star}$ and SFR}
\label{sed}

We measure M$_{\star}$ and SFR for each galaxy from fitting the full SED produced from all available multi-wavelength data.
The knowledge of accurate spectroscopic redshifts is a key advantage in the SED fitting process as it minimizes possible degeneracies occurring when trying to measure both a photometric redshift and a set of physical parameters from the same SED fitting process.
We therefore perform the SED fitting for each galaxy using the spectroscopic redshifts of our sample.
Spectral energy distribution (SED) fitting is performed using the code Le Phare \citep{Arnouts1999,Ilbert:06}. 
The core engine for Le Phare is template fitting to the photometric dataset of each galaxy using a range of templates coming from \citet[][,hereafter BC03]{Bruzual:03} models  and is using a \citet{Chabrier:03} IMF. 
We use exponentially declining star–formation histories $SFR \propto e^{-t/\tau}$ ($\tau$ in the range 0.1 Gyr to 30Gyr), and
two delayed SFH models with peaks at 1 and 3Gyr.
The SEDs are generated for a grid of 51 ages (in the range 0.1 Gyr to 14.5 Gyr).
A \citet{Calzetti:00} dust extinction law was applied to the templates with E(B$-$V) in the range 0 to 0.5. 
We used models with two different metallicities. 
The best fit is obtained by means of the best $\chi^2$ between the observed SED and the set of templates. 
A key feature of Le Phare is the realistic addition of emission lines to the templates, as extensively described in \citet{Ilbert:09}. In short this is performed using the star formation rate of each template: the SFR is transformed into line equivalent widths using a standard set of transformations issued from case B line recombination and these lines are then added to the stellar population models. A constant ratio is set between emission lines (before correcting them for extinction), and the flux of the emission lines is allowed to vary within a factor 2.
In the redshift range considered here several emission lines can reach high enough equivalent widths to significantly modify near--IR broad band magnitudes, most notably $H\beta$-4861\AA, the [OIII]4959--5007\AA ~doublet and the $H\alpha$--6562\AA ~line. When these lines are in emission they can change the magnitude in one of the redder bands of our photometric dataset, the K--band for z$\simeq 3$ up to the IRAC 3.6 and 4.5$\mu$m bands for $z \sim 4$ to $6$, altering the computation of e.g. stellar masses and SFR if the SED fitting is using only stellar continuum templates. 
In the redshift range $2<z<6$ the K band and IRAC 3.6 and 4.5$\mu$m bands cover a wavelength from $\sim 4000$\AA ~to 1$\mu$m, a most important domain to derive the stellar mass, as M$_{\star}$ is roughly proportional to the observed flux at these rest wavelengths in the SED fitting process. 
Including emission lines may change the observed flux in the affected bands by more than one magnitude, which in turn may change a SED--derived stellar mass by $0.1-0.2$ dex compared to standard SED fitting without emission lines included \citep[see e.g.][]{Salmon:14,deBarros:14}.
This was originally discussed in \citet{Ilbert:09}, and further studied recently \citep{Stark:13,deBarros:14,Salmon:14} when it was realized that emission lines may significantly bias  SFR and sSFR measurements at redshifts $z>2$ if not properly taken into account in the SED fitting.
The process of adding emission lines to templates is likely to be not fully controlled beyond the first principles, because the exact strength of each line is not known a priori for each galaxy, and it may therefore introduce some uncertainties leading to a larger dispersion in the distribution of measured parameters. 
One important  point is the dispersion in the relation between line strength and SFR, and the evolution of this relation with redshift. 
While recent studies have provided some complex means to take into account the main emission lines \citep[e.g.][]{deBarros:14}, the simple approach used by Le Phare significantly limits the sensitivity of the derived stellar mass and SFR to the presence of emission lines in the most important rest--frame bands. 


\section{The evolution of the SFR--M$_{\star}$ relation}
\label{sfr_mass}

\subsection{The star--forming main sequence up to z$\sim5$}
\label{ms}

As a unique feature, the VUDS survey covers the whole redshift range from the local universe up to z=5.5 using spectroscopically confirmed galaxies. The derivation of the SFR and M$_{\star}$ parameters is done following the same methods and input data, making the relative comparison of the SFR--M$_{\star}$ relations at different redshifts less prone to systematics.  

\begin{figure*}
   \centering
   \includegraphics[width=14cm]{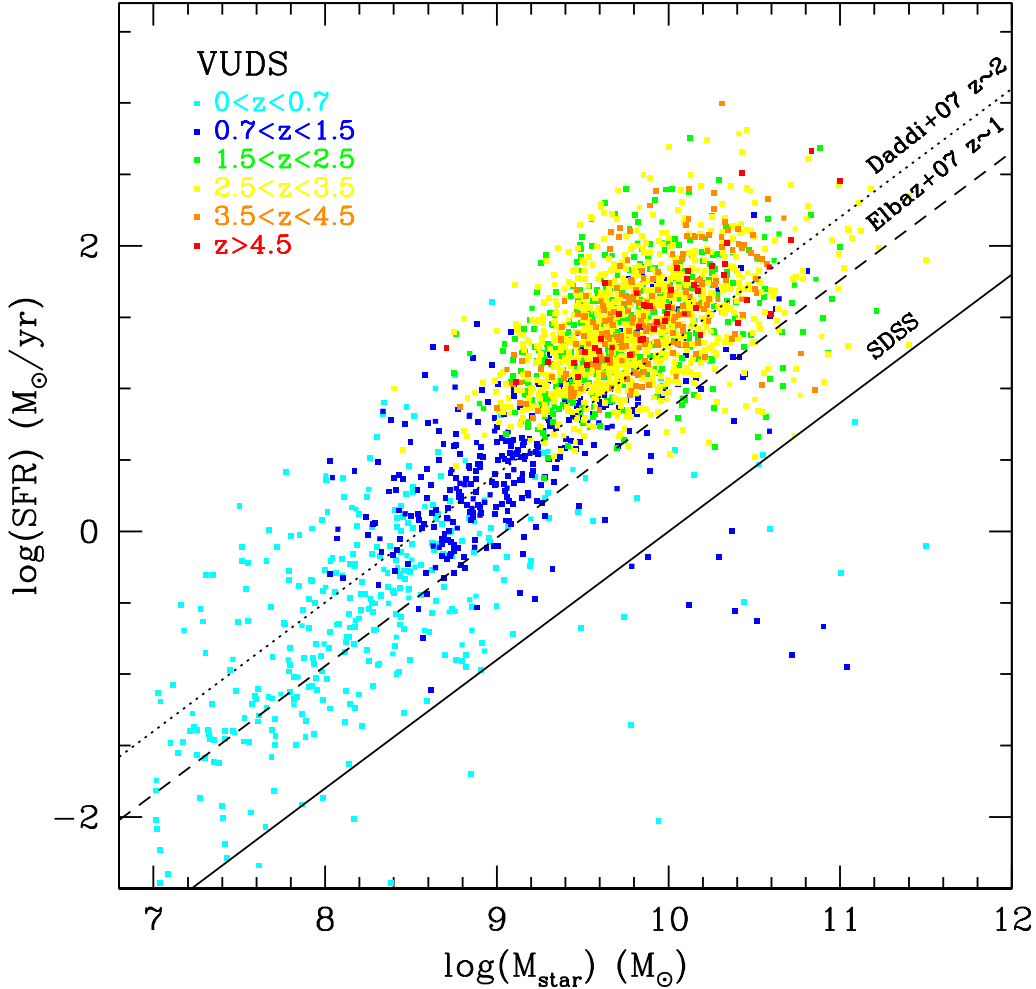}
      \caption{The SFR--M$_{\star}$ relation for VUDS star--forming galaxies. Points are color--coded depending on the spectroscopic redshift of each galaxy as indicated in the inset. 
A clear evolution of the SFG main sequence is observed in the VUDS sample up to the highest redshifts z$\simeq$5.
This is confirmed to z$\sim$2 when comparing VUDS data to the main sequence measured from the SDSS at z$\sim$0.2 (Peng et al. 2010; full line), the MS at z$\sim$1 by Elbaz et al. (2007; dashed line), and the MS of Daddi et al. (2007; dotted line) at z$\sim$2.
At $z>2$ the VUDS data appear to lie significantly above the \citet{Daddi:07} main sequence.}
         \label{sfr_all}
   \end{figure*}

We plot the SFR--M$_{\star}$ relation for all VUDS galaxies in Figure \ref{sfr_all}, color--coded as a function of redshift. 
It is immediately visible that the distribution of VUDS galaxies over this large redshift range does no follow a single main sequence relation. 
On average our sample galaxies are more than 1 dex above the local main sequence (MS) of star forming galaxies in the SDSS \citep{Peng:10} at any redshift considered in this study. 
A significant fraction of our data also lie above the \citet{Daddi:07} MS at z$\sim$2. 

\begin{figure*}
   \centering
   \includegraphics[width=18cm,bbllx=1,bblly=200,bburx=591,bbury=620,clip=]{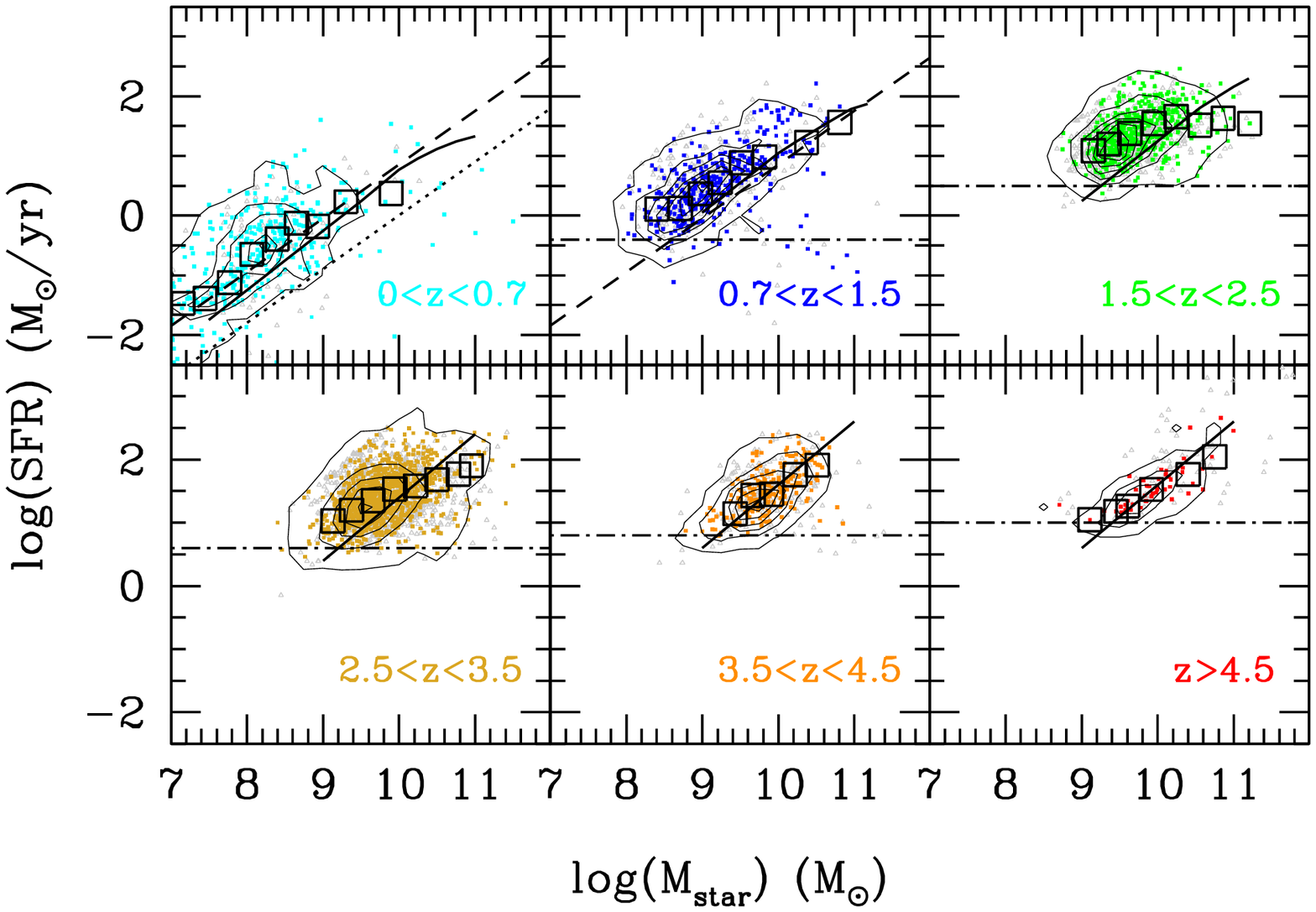}
      \caption{The SFR--M$_{\star}$ relation for VUDS star--forming galaxies per redshift bins from z$\sim$0.5 to z$\sim$5. 
In each panel coloured data points represent individual galaxies while contours show the density of galaxies.
VUDS galaxies with redshifts $\sim100$\% reliable (flags 3 and 4, see Le F\`evre et al. 2014) are plotted with filled (coloured) heavy symbols, while galaxies with redshifts $\sim70-75$\% reliable (flag 2) are indicated with light grey open symbols. 
The black squares in each panel represent the median SFR in increasing mass bins obtained from the individual galaxies.
The main sequence observed locally by the SDSS \citep{Peng:10} and at z$\sim$1 by \citet{Elbaz:07} is indicated in the 0$<$z$<$0.7 panel by the dotted and dashed lines, respectively. 
At higher redshifts the contiguous line indicates the MS of \citet{Schreiber:14}, including the high mass turn--off observed in their study, and the \citet{Elbaz:07} $z=1$ relation is added as the dashed line in the $0.7<z<1.5$ panel.
Horizontal dot--dashed lines in each panel indicate the limit in SFR above which our data are statistically complete, as imposed by the $i_{AB} \leq 25$ limit of our sample.}
         \label{sfr_z}
   \end{figure*}

The evolution of the MS location with redshift is best seen when plotting the SFR vs. M$_{\star}$ in several redshift bins as presented in Figure \ref{sfr_z} with single galaxy points as well as median values in stellar mass bins. In the lowest redshift bin of our sample at z$<$0.7 the galaxies with very small masses down to M$_{\star}$$\sim$$10^{7}$ M$_{\sun}$ lie quite high in SFR, on the MS of z$\sim$1 \citet{Elbaz:07}, while intermediate mass galaxies 8.5$<log($M$_{\star})$$<$9.5 are in between the MS from SDSS \citep{Brinchmann:04,Peng:10} and the z$\sim$1 MS  of \citet{Elbaz:07}, as expected in this redshift bin. 
At z$\sim$1 our data are slightly above in SFR (or less massive) than in the \citet{Elbaz:07} relation. 
The difference between our data and data in the literature is $\sim$0.1dex and could be the result of different systematics between our study and other studies in the literature.
Going to higher redshifts where the bulk of VUDS galaxies are identified, we find that our galaxies are reaching strong star formation rates SFR$>$100 M$_{\sun}$/yr, with the sample still containing high mass galaxies up to a few $10^{11}$M$_{\sun}$, thanks to the large volume covered. 
In the redshift bin z=[1.5,2.5] VUDS galaxies with $log($M$_{\star})$$<$10.25 are above the $z\sim2$ MS relation of \citet{Schreiber:14} likely partly due to the median redshift of our data $z=2.37$ (resulting from the photometric redshift selection of the VUDS sample). 
For z=[2.5,3.5] lower mass galaxies are also above the  \citet{Schreiber:14} relation, while at z$>$3.5 the VUDS data are quite well centered on this relation up to z$\sim$5.5.
Over the redshift range z=[1.5,3.5] it can be clearly seen that a significant fraction the most massive galaxies in our sample with $log$(M$_{\star}$)$>$10.25 are below the \citet{Schreiber:14} relation. This is further discussed in the next section. 

Prior to discussing the SFR--M$_{\star}$ relation it is important to take into account that the VUDS selection function includes a $22.5 \leq i_{AB} \leq 25$ magnitude selection which implies a low SFR limit, in effect a Malmquist bias evolving with redshift \citep[see e.g.][]{Reddy:12}. 
The high magnitude cutoff $i_{AB}$=22.5 limits the detection of massive and star--forming galaxies at z$<\sim$1 but it is not expected to exclude any massive galaxies at higher redshifts as was verified from the VVDS survey \citep{LeFevre:13a}.  
We use the semi--analytic model of \citet{Wang:08} applied to the COSMOS field to better identify the statistical completeness of our data with this selection limit (here we use the term {\it statistical completeness} as the ability to identify galaxies with a certain property in the VUDS fields). Applying the VUDS magnitude selection to the simulation we find that the magnitude limit restricts the sample to galaxies with higher SFR as redshift increases, as indicated in Figure \ref{sfr_z}. VUDS is essentially statistically complete in the SFR-M$_{\star}$ plane up to M$_{\star} \sim 5 \times 10^{9}$ M$_{\sun}$ and $-2<log(SFR)<1.8$ for $z<0.7$. It is statistically complete for $log(SFR)>-0.4, 0.5, 0.6, 0.8$ and $1$ at $z \sim$ 1, 2, 3, 4, and 5, respectively.  
We further note that if trying to fit a linear main sequence $log(SFR) = log(M_{\star})^{\alpha}$ the slope $\alpha$ would then be artificially flattened at lower masses by the magnitude selection, and we do not attempt in this paper to quantify the MS slope at stellar masses for which this effect is at work. 

To further analyse the behaviour of the SFR as a function of mass at different epochs, we compute median SFRs in increasing mass bins, imposing the SFR limits quoted above, as plotted in Figure \ref{sfr_z}. 
From the median values we see departure from a linear main sequence relation at both the lowest and the highest masses. 
These observed trends are the consequence of two effects: the bias against low SFR and low M$_{\star}$ galaxies due to the VUDS selection function as described above, and a turn--off of the SFR--M$_{\star}$ relation at high masses, as is further discussed in Section \ref{turnoff} below.

\subsection{A turn--off in the SFR--M$_{\star}$ relation: evolution with redshift}
\label{turnoff}

As the VUDS survey is conducted in a 1 square degree area, it is picking up more of the rarer galaxies at the high mass end of the M$_{\star}$ distribution. 
This is a key advantage compared to most other surveys probing smaller areas, and allows to study the behaviour of the SFR--M$_{\star}$ relation reliably at the highest masses. 

We observe a significant departure from a linear main sequence at high masses and at all redshifts $z<3.5$ probed by this study, which seems to indicate a physical dependency of the SFR on M$_{\star}$. 
In the redshift bin $0.7<z<1.5$, our data indicate a turn--off at a  mass M$_{\star} \sim  10^{10}$ M$_{\sun}$. 
Going to higher redshifts, we observe a significant turn--off of the SFR--M$_{\star}$ relation in both of the $1.5<z<2.5$ and $2.5<z<3.5$ redshift bins, occurring at M$_{\star} \sim 1.5$ and $\sim 2.5 \times 10^{10}$ M$_{\sun}$, respectively. 
In $1.5<z<2.5$ our median SFR measurements stay almost constant from log(M$_{\star}$)=10.2 to 11.2 at log(SFR)=1.6$\pm0.1$, hence presenting a deficit in SFR by $\sim1$dex compared to the SFR expected extrapolating the SFR at log(M$_{\star}$)=10.2 if SFR$\propto$M$_{\star}$. 
Similarly in $2.5<z<3.5$, extrapolating from log(M$_{\star}$)=10.2 would lead to log(SFR)=2.5 while we measure log(SFR)=1.9$\pm0.1$. 
In the redshift bin $3.5<z<4.5$ a turn--off may be happening at around M$_{\star} \sim 3 \times 10^{10}$ M$_{\sun}$ but is not significant at less than $1.5\sigma$. 
At z$\sim$5 we do not observe a turn--off and the SFR--M$_{\star}$ relation seems to be linear over the mass range probed except at the low--mass end as expected due to the sample selection function as described in Section \ref{ms}. 

The VUDS data selection ensures that most of the strongest SFGs at high masses are included in the sample, unless there exists a substantial population of high mass high SFR heavily obscured galaxies at $z>1.5$. 
If any bias is present at high masses in VUDS, it would be against a population of low star forming galaxies with $SFR<3 M_{\sun}$/yr at $z>\sim2.5$. 
Such a population has been identified at z$\sim$2-3 as claimed by \citet{Whitaker:13}, and remains elusive at $z>3$. 
A large low SFR population at these redshifts might indicate galaxies which have already quenched pushing the onset of star formation at even higher redshifts, and it remains possible that a low SFR and high M$_{\star}$ population could exist in small numbers at these redshifts. 
However, such a population would only lower the average SFR at high M$_{\star}$ in the SFR--M$_{\star}$ relation and therefore further amplify the turn--off in the main sequence reported here, and our results remain qualitatively robust to any putative low SFR -- high M$_{\star}$ population. 
What we observe is a small population of galaxies with intermediate SFR, which might represent galaxies on their way to the passive population observed at lower redshifts. 
While these objects are star--forming with $SFR>3 M_{\sun}$/yr at $z>\sim2.5$ and cannot be excluded from the analysis of a star-forming population, they are definitely off the MS extrapolation to high mass and are driving median SFR values down. 
This population deserves a closer look \citep{Tasca:14c}. 
Another population which is likely missed by VUDS is galaxies which are heavily star--forming but strongly obscured by dust. 
As shown by \citet{Rodighiero:14} and \citet{Whitaker:12} comparing UV--selected and IR--selected samples, we expect to have missed less than $7\%$ of high--mass high--SFR galaxies lying mostly above the MS. This possibly missed fraction is not sufficiently high to produce the trend we observe in our data.

Departure from a linear MS have been reported in the literature. 
From a radio stacking analysis, \citet{Karim:11} found tentative evidence for curvature of the star formation sequence. 
Up to $z \sim 2.5$ our data are in agreement with the results by \citet{Whitaker:12} who find that the SFR--M$_{\star}$ relation follows a power--law $SFR \propto M_{\star}^{0.6}$ rather than a linear relation. 
\citet{Whitaker:14} bring further evidence for a mass--dependent behaviour of the SFR--M$_{\star}$ relation with a steep slope for low--mass galaxies, and a shallower slope at high mass from the 3D--HST survey using photometry from CANDELS. 
Furthermore our results identify for the first time a similar behaviour at $z>2.5$. 
We find that for massive galaxies with M$_{\star} > 10^{10}$ M$_{\sun}$ the rise in median SFR with mass is slower than for galaxies at lower masses, and this property seems to hold to z$\sim$3.5. 
To understand the behaviour at the massive end, Whitaker et al. (2012, 2014) compare UV--selected and LIR selected samples and conclude that the linear MS relation observed from UV rest--frame data is the result of the selection function truncating galaxies with high stellar mass and high SFR but with a lot of dust extinction, and that without the UV selection the relation would show downward curvature. 
The VUDS selection function in effect is UV--selected because of the $i_{AB}\leq25$ cutoff but we do not see a continuous linear SFR--M$_{\star}$ relation but we rather observe a high--mass turn--off. 
We therefore infer that the observed turn--off is not only related to dust--obscured galaxies, but also to a general lack of strongly star--forming galaxies at high masses, at least in the redshift range $1<z<3.5$. 

A departure from a linear MS relation from lack of strongly star--forming galaxies can be interpreted as the result of star--formation quenching.
Quenching could be produced either because the gas supply is reduced, e.g. if the rate of gas accretion is reduced, or because stars cannot form as efficiently, e.g. because of feedback or environment effects. Galaxy mergers would have the effect of bringing galaxies out of the MS even if the SFR would increase during the merger event \citep{Peng:10}.  
Interestingly we find that the turn--off mass where departure from a linear main sequence seems to occur is decreasing with decreasing redshift, going from M$_{\star} \sim 2.5 \times 10^{10}$ M$_{\sun}$ at z$\sim$3 to M$_{\star} \sim 10^{10}$ M$_{\sun}$ at z$\sim$1 and M$_{\star} \sim 8 \times 10^{8}$ M$_{\sun}$ at $z \sim 0.4$. 
This downsizing trend echoes downsizing in other properties \citep{Cowie:96} and is expected in models relating quenching to the fast evolution of the star formation rate density \citep[e.g.][]{Peng:10}. 

At the highest redshifts of our sample $z>4.5$, we find that the SFR follows a linear $SFR \propto M_{\star}^{\alpha}$, without an apparent turn--off at high mass.
Although our sample may not be large enough to identify a turn--off, this result is comparable to \citet{Steinhardt:14}. 
Assuming that the fraction of high mass--high SFR dust--obscured galaxies is not higher at these redshifts than for $z<4.5$, a possible interpretation is that at these high redshifts the SFR quenching mechanisms are not yet fully in place. 
As the SFRD is in a strongly increasing phase \citep{Bouwens:14}, the number of SNe capable to drive strong winds may not yet be sufficient for feedback to be strong enough to lower star formation. Similarly, the number and mass of central massive black holes may not bring AGN feedback to sufficient levels to quench star formation. 
From our data we therefore argue that star formation quenching mechanisms may become most efficient starting at $z<\sim4$. 
The downsizing in the mass turn-off may further indicate that this quenching progresses steadily to lower masses as redshift decreases.


\section{The evolution of the specific star formation rate since $z \simeq 5$}
\label{ssfr_z}

In this section we explore the evolution with redshift of the median value of the sSFR. 

\begin{figure*}
   \centering
   \includegraphics[width=18cm,bbllx=1,bblly=210,bburx=591,bbury=630,clip=]{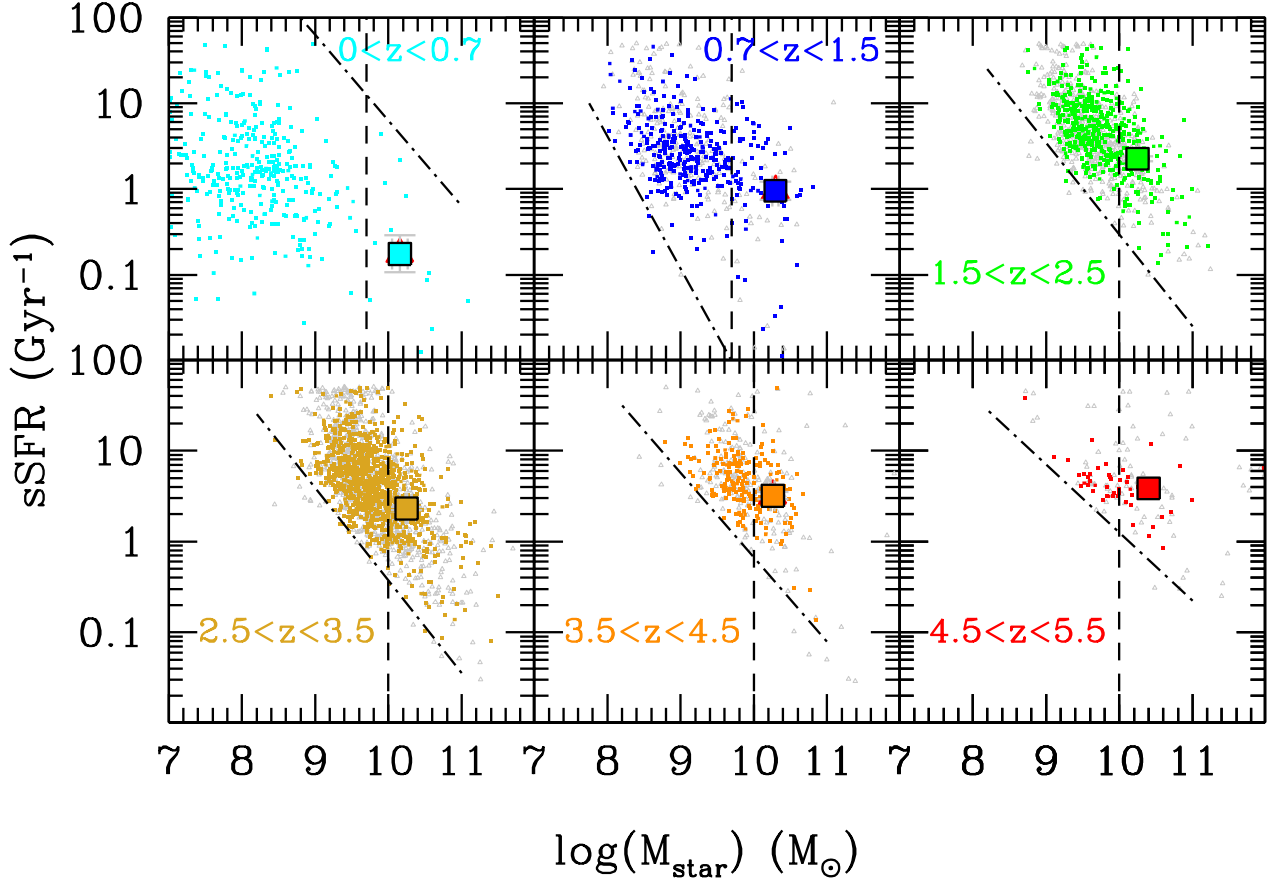}
      \caption{The distribution of sSFR vs. M$_{\star}$ for VUDS galaxies in several redshift bins with median ranging from z$\sim$0.4 to z$\sim$5. 
VUDS galaxies with redshifts $\sim100$\% reliable (flags 3 and 4, see Le F\`evre et al. 2014) are plotted with filled (coloured) heavy symbols, while galaxies with redshifts $\sim70-75$\% reliable (flag 2) are indicated with light grey open symbols. 
The stellar mass limit above which the median sSFR is computed is indicated by the vertical dashed lines in this panel, chosen to be M$_{\star} \geq 5 \times 10^{9} M_{\sun}$ for $z \leq 1.5$ and M$_{\star} \geq 10^{10}$ M$_{\sun}$ for $z>1.5$.
The limit in the data as imposed by the VUDS selection function is indicated by the dot--dash line in each panel, as validated imposing a similar selection function on the \citet{Wang:08} SAM. 
The median sSFRs above the mass limits are indicated by the large black colour--shaded squares. 
The error bars represent the $1 \sigma/\sqrt(N)$ error on the median from the  $1 \sigma$ dispersion in the data and the N galaxies in the bin; they are lower than the size of the data points, except for $z<0.7$.
      }
         \label{ssfr_red}
   \end{figure*}

The distribution of sSFR as a function of stellar mass is shown in Figure \ref{ssfr_red} for different redshift bins. 
The VUDS selection function implies some restrictions in probing the sSFR-M$_{\star}$ plane, and we plot the empirical limits on the sSFR--M$_{\star}$ resulting from the VUDS selection  in Figure \ref{ssfr_red}.
Below $z=1.5$ the VUDS sample is statistically complete in sSFR and mass above M$_{\star}= 5 \times 10^9$ M$_{\sun}$. 
Above $z=1.5$ we use a low mass cutoff of M$_{\star} = 10^{10}$ M$_{\sun}$ to compute the median sSFR. 
Above these mass limits, we find a large range in sSFR ranging more than 2 dex: in z=[1.5,3.5] we observe sSFR as low as $\sim$0.3, and going up to $\sim$30.  At z$>$3.5 the lowest sSFR are $\sim$1, and go beyond $\sim$30. 
The median value in sSFR and M$_{\star}$ is indicated in each panel, and discussed below. 
The sSFR decreases with M$_{\star}$ as expected from the lower SFR for high M$_{\star}$ (see Section \ref{sfr_mass}, and Whitaker et al. 2012, 2014). 
It is important to note that because of the large 1 deg$^2$ field and corresponding large volume surveyed by VUDS compared to the smaller $\sim$170 arcmin$^2$ of CANDELS \citep{Salmon:14}, $\sim$55 arcmin$^2$ of \citet{Gonzalez:14} in ERS and HUDF, or $\sim$300 arcmin$^2$ of \citet{Stark:13} in GOODS, our sample includes a larger number of galaxies with high masses M$_{\star} > 10^{10}$M$_{\sun}$. 
Samples in smaller fields than explored in VUDS are likely missing the highest mass galaxies and therefore may not sample enough galaxies to identify the high mass behaviour of the sSFR. 

\begin{figure*}
   \centering
   \includegraphics[width=14cm]{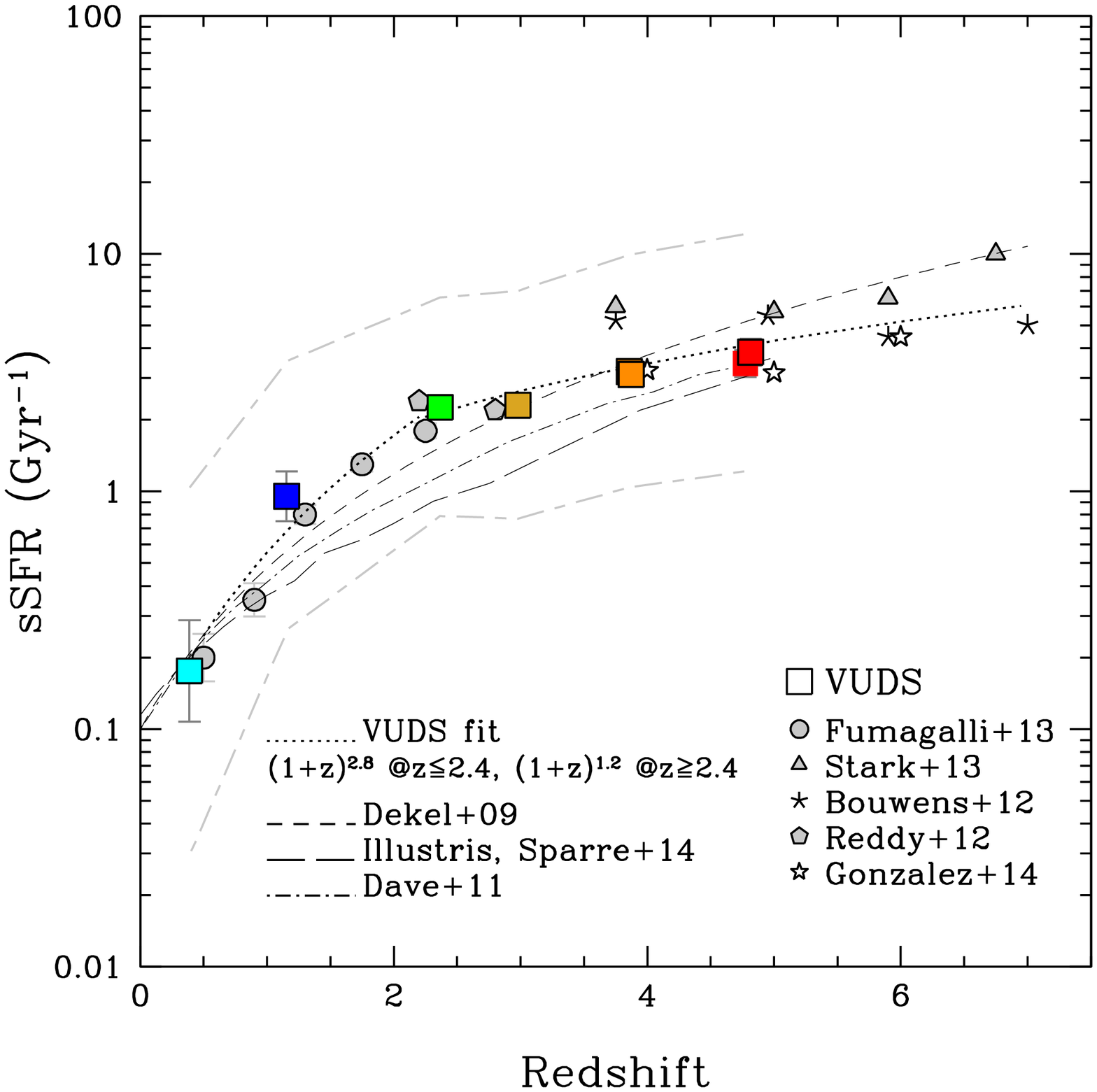}
      \caption{The evolution of the sSFR with redshift for VUDS star forming galaxies, obtained computing
               the median sSFR value for M$_{\star} \geq 5 \times 10^{9} M_{\sun}$ when $z \leq 1.5$ 
               and M$_{\star} \geq 10^{10}$ M$_{\sun}$ for $z>1.5$.
	       Error bars on each VUDS data point indicate the $1\sigma$ error on the median of the observed sSFR distribution and
               are generally smaller than the size of the data points. 
               For the points at $z=3.9$ and $z=4.8$ flag 2, 3, 4 and 9 have been included to increase
               the sample size, and we also indicate the median values for flag 3 and 4 only, represented by the
               coloured square symbols which lie at only slightly lower values (almost undistinguishable for $z=3.9$).
               At $z<2.4$ we find that the sSFR evolution follows $(1+z)^{2.8}$, while the evolution is slower at $2.4<z<5$
               following $(1+z)^{1.2}$ (dotted line).
	       The $\pm1\sigma$ of the sSFR distribution is represented by the light grey short dash -- long dash lines below and above the median values.
               Several other data sets from the literature are plotted as discussed in the text
               (Reddy et al. 2012, Bouwens et al. 2012, SFGs from Fumagalli et al. 2013, Stark et al. 2013, Gonzalez et al. 2014); 
               some of these points have been
               slightly shifted to avoid overlap with the VUDS data points, see the exact values in these papers.
               Several models predicting the evolution of the sSFR are indicated, including galaxy growth dominated
               by cold gas accretion (Dekel et al. 2009, dashed line, normalized to sSFR(z=0)=0.1), 
               the hydrodynamical simulation of Dav\'e et al. (2011, dot-dashed line),
               and the latest results from the Illustris hydrodynamical simulation (Sparre et al. 2014, long-dashed line). 
              }
         \label{ssfr_evol}
   \end{figure*}

The evolution of the median sSFR with redshift is presented in Figure \ref{ssfr_evol}, and median sSFR measurements are listed in Table \ref{ssfr_table}. 
We compute the error on the median value as $1\sigma/\sqrt(N_{obj})$ where $\sigma$ is the standard deviation in the sSFR distribution and $N_{obj}$ is the number of galaxies in the redshift bin considered.
We compute the median sSFR above a stellar mass lower limit of M$_{\star} = 5 \times 10^{9}$ M$_{\sun}$ for $0<z<1.5$ and M$_{\star} \geq 10^{10}$ M$_{\sun}$ for $z>1.5$. 
At $z<1.5$ our data are fully consistent with the results presented in \citet{Fumagalli:13} from the 3D--HST survey,
and somewhat higher than \citet{Ilbert2014} likely due to the difference in the mass range sampled. 
At redshifts z$\sim$2$-$3 our data is in excellent agreement with the data presented in \citet{Reddy:12} using BM, BX, and LBG galaxies \citep{Steidel:03}.
At redshifts z$\sim$3$-$5 our median sSFR measurements compare well with \citet{Gonzalez:14} but are a factor $\sim$1.4 lower than \citet{Stark:13} or \citep{Bouwens:12}. 

From our data we observe a strong evolution of the median sSFR from z$\sim$0.4 to z$\sim$2.3: the observed sSFR evolution in the VUDS dataset is very steep, decreasing by a factor $\sim13$ from $z=2.3$ to $z=0.4$. 
At z$\sim$2.3 we find a median sSFR  $sSFR(z=2.3)=2.3\pm0.16$ Gyr$^{-1}$. 
Parametrizing the sSFR evolution as $log(sSFR_z) = A + \Phi \times log(1+z)$ we find $A=-10.1\pm0.03$ and $\Phi=2.8\pm0.2$ for z$<$2.3, somewhat less steep than discussed by \citet{Fumagalli:13} who report a slope of $\Phi\sim$3. 
At $z>2.3$ the sSFR continues rising reaching $sSFR(z=4.8)=3.9\pm0.5$ Gyr$^{-1}$. 
We find that in our data the evolution between z$\sim$5 and z$\sim$2 is slower than for z$<$2, and is best parametrized with $A=-9.3\pm0.02$ and $\Phi=1.2\pm0.1$.
This trend does not change if we impose a lower cut in sSFR (e.g. log(sSFR)$>$-9.2 and -9 in z=[2.5,3.5] and z=[3.5,4.5], respectively). 
This is further discussed in Section \ref{discuss}.

\begin{table*}
\caption{Median redshifts and specific star formation rates from the VUDS sample in $0<z<5$.  }
\label{ssfr_table}    
\begin{tabular}{c c c c c c c}       
\hline\hline                
z--range & Median &  \multicolumn{2}{c}{N$_{galaxies}$} & M$_{\star}$  &  log(sSFR)  &   Error on mean sSFR  \\
        &  z     &   Total  & Above mass cut       & & \multicolumn{2}{c}{sSFR in yr$^{-1}$}      \\
\hline                       
0$-$0.7$^1$   &       0.39   & 350 & 13   &    10.16   &   -9.755  &    0.21     \\
0.7$-$1.5$^1$ &       1.15   & 364 & 29   &    10.29   &   -9.020  &    0.11     \\
1.5$-$2.5$^1$ &       2.37   & 461 & 217  &    10.25   &   -8.646  &    0.03     \\
2.5$-$3.5$^1$ &       2.98   & 977 & 561  &    10.25   &   -8.641  &    0.02     \\
3.5$-$4.5$^1$ &       3.86   & 194 & 79   &    10.26   &   -8.519  &    0.06     \\
3.5$-$4.5$^2$ &       3.87   & 218 & 154  &    10.22   &   -8.507  &    0.04     \\
4.5$-$5.5$^1$ &       4.82   & 50  & 22   &    10.40   &   -8.414  &    0.11     \\
4.5$-$5.5$^2$ &       4.78   & 85  & 77   &    10.33   &   -8.461  &    0.06     \\
\hline                                 
\end{tabular}
 \begin{list}{}{}
 \item[$^{\mathrm{1}}$] Values are computed for flags 3,4
 \item[$^{\mathrm{2}}$] Values are for flags 2,3,4,9.
 \end{list}
\end{table*}

\section{Discussion and summary}
\label{discuss}

We use a spectroscopic sample of 2435 star--forming galaxies with highly reliable  spectroscopic redshifts (flag 3 and 4) from the VUDS survey to study  the evolution with redshift of the SFR--M$_{\star}$ relation and of the sSFR, up to a redshift z$\sim$5. 
We use an additional sample of 2096 galaxies with reliable spectroscopic redshifts (flag 2 and 9) to consolidate statistical analysis when necessary. 
SED fitting using the code Le Phare is performed on the extensive photometric data at the spectroscopic redshift of each galaxy and taking into account the contributions from nebular emission lines.  
The knowledge of the spectroscopic redshift enables to limit degeneracies in computing the SFR and M$_{\star}$. 
Our data cover a range of M$_{\star}$ from $10^9$ to $10^{11}$ M$_{\sun}$ at z$\sim$2, as enabled by the large 1 square degree field surveyed, while at our highest redshifts $4.5 < z < 5.5$ we observe galaxies with M$_{\star}>10^{9.4}$ M$_{\sun}$. 
We then discuss the observed SFR--M$_{\star}$ relations as well as the evolution of the sSFR with redshift. 
The VUDS data used in this study cover a large redshift range $0<z<\sim5$ with a large number of galaxies at $z>2$, which allows for the first time a consistent study of evolution from a single survey with the same selection function, avoiding the difficulties in comparing inhomogeneous samples.  

The SFR--M$_{\star}$ relation strongly evolves with redshift. 
We clearly identify a main sequence along which galaxies lie, and the position of the main sequence evolves with redshift.
We observe that a main sequence holds above z$\sim$2 and up to the  highest redshifts z$\sim5$ in our sample, as observed in other datasets \citep[e.g.][]{Stark:13,Salmon:14}.  
We find that for redshifts $z<3.5$ the SFR--M$_{\star}$ relation at high masses is not a linear extrapolation of the relation at the lower masses in agreement with Whitaker et al. (2012, 2014). Furthermore we find that the mass at which the main sequence becomes non--linear decreases with decreasing redshift from  M$_{\star}\sim 2.5 \times 10^{10}$ M$_{\sun}$ at z$\sim$3 down to M$_{\star}\sim 10^{10}$ M$_{\sun}$ at z$\sim$1 and M$_{\star}\sim$8$\times 10^{8}$ M$_{\sun}$ at z$\sim$0.4. 
Interpreting the turn--off at high masses as the effect of star formation quenching, this downsizing pattern may indicate that this quenching is gradually progressing from high masses at high redshifts to low masses at low redshifts.
At $z>3.5$ the main sequence seems to become more linear. 
This may indicate that quenching processes are not yet fully active at these redshifts, and that possible quenching processes like SNe or AGN feedback would only bring--up sufficient energy release to significantly quench star--formation at redshifts below z$\sim$3.5.

We compute the sSFR from z$\sim$0.1 all the way up to z$\sim$5. 
We find that the dependence on redshift of the median sSFR is different at low redshifts (z$<$2.3) or at high redshifts 2.5$<$z$<$5. 
At the lower redshifts, we observe a strong rise in sSFR by a factor $\sim13$ from $z=0.4$ to $z=2.3$ reaching a $sSFR(z=2.3)=2.3 Gyr^{-1}$ following an evolution $\propto(1+z)^{\Phi}$ with $\Phi=2.8\pm0.2$ similar to that reported in the literature \citep[e.g. SFGs in][]{Fumagalli:13}. 
From $z=2.3$ to $z=4.8$ the sSFR continues to increase but at a slower rate. 
At redshifts $2<z<3$ our data are in excellent agreement with \citet{Reddy:12}.
At the highest redshift end of our dataset $3<z<5$ our data are broadly consistent with data in the literature \citep{Bouwens:12,Stark:13,Gonzalez:14} when considering systematics in computing the sSFR. 
At a redshift z$\sim$4 we find that the VUDS median sSFR is the same as \citet{Gonzalez:14}, $\sim$1.7$\times$ lower than \citet{Bouwens:12}, and $\sim$2$\times$ lower than \citet{Stark:13}. 
Our highest redshift measurement at z$\sim$5 is about 20\% higher than \citet{Gonzalez:14} and $\sim30$\% lower than \citet{Bouwens:12} or \citet{Stark:13}. 
From $z=2$ to $z=5$ we find $sSFR \propto (1+z)^{\Phi}$ with $\Phi=1.2\pm0.1$ for galaxies selected above a mass limit M$_{\star} = 10^{10}$ M$_{\sun}$. 
The observed evolution at $z>2.3$ in our data is somewhat flatter than the evolution $\propto(1+z)^{3.4}$ as reported in \citep{Salmon:14} but for galaxies selected at a smaller mass M$_{\star} = 10^{9}$ M$_{\sun}$ $\sim0.5$dex lower than ours. 

The flattening of the sSFR evolution beyond redshift z$\sim$2 in our data is compared to models in Figure \ref{ssfr_evol}. 
Models based on cold accretion--driven galaxy growth are expected to follow  sSFR $\propto (1+z)^{\Phi}$ with $\Phi=2.25$  \citep{Dekel2009}. 
The sSFR at a given mass in these models is lower than our data and other data in the literature for z$<\sim$2.  
The VUDS data as well as other observational results are located at significantly higher sSFR than hydrodynamical simulations incorporating parametrised galactic outflows \citep{Dave:11}, as well as above the latest Illustris hydrodynamical moving--mesh simulation \citep{Sparre:14}, as drawn on Figure \ref{ssfr_evol}. 
At higher redshifts these hydrodynamical simulations intersect the observed data at z$\sim$4 but with a steeper slope than in observational data. 
The comparison of observed data with current models therefore seems to indicate that the sSFR evolution does not follow a pure  accretion driven galaxy mass growth. 

Several important physical processes are known to be at play which could well alter the simplified cold-accretion growth picture. We presented observational evidence in Section \ref{turnoff} supporting a picture where star formation quenching starts to be efficient below z$\sim$3.5, in effect reducing the increase in SFR expected from cold accretion alone. Mergers are ubiquitous at all redshifts, reaching a major merger fraction of $\sim20$\% at z$\sim$1.5 (L\'opez-Sanjuan et al. 2013) and staying high to beyond z$\sim$3 \citep{Tasca:14} driving the mass growth in a different way than cold gas accretion does. 
The effect of merging on the SFR vs. M$_{\star}$ relation can be viewed as a straight shift in M$_{\star}$ at fixed SFR, with equal mass (major) mergers  doubling M$_{\star}$. 
Repeated minor merger events would also participate to this trend of increasing M$_{\star}$ with only a modest increase in SFR if the merging galaxies are of lower SFR than the primary galaxy, leading to a further flattening of the sSFR evolution with redshift. Mergers in essence produce a mass increase from stars formed beyond the immediate environment of the galaxy adding to the stars formed in the galaxy breaking the SFR$\propto$M$_{\star}$ relation. 
It is then likely that these processes combine with cold gas accretion to modulate the SFR and mass growth to produce the observed sSFR evolution.

Despite remarkable improvements in the observational data, the main limitation to the study of the sSFR remains the uncertainties and systematics errors associated to the computation of SFR and M$_{\star}$. 
Progress in measuring SFR on individual galaxies beyond z$\sim$2.5 will require a significant improvement in observing capabilities such as what is expected from the JWST, able to follow direct SFR tracers like the $H\alpha$ line to z$\sim$6.6 and beyond. 
Improvements on M$_{\star}$ estimates will be even harder to get because of current unknowns on the IMF evolution with redshift or more complex SFH than in current models. 
It is nevertheless clear, and perhaps not surprising, that the sSFR now firming up from the VUDS observational results presented in this paper and others in the literature requires models with a more balanced mix of physical processes than models dominated by cold gas accretion developed up to now. 
These new generation observations and models will help in turn to better understand the star formation history and galaxy stellar mass assembly.


\begin{acknowledgements}
We thank ESO staff for their continuous support for the VUDS survey, particularly the Paranal staff conducting the observations and Marina Rejkuba and the ESO user support group in Garching.
This work is supported by funding from the European Research Council Advanced Grant ERC--2010--AdG--268107--EARLY and by INAF Grants PRIN 2010, PRIN 2012 and PICS 2013. 
AC, OC, MT and VS acknowledge the grant MIUR PRIN 2010--2011.  
This work is based on data products made available at the CESAM data center, Laboratoire d'Astrophysique de Marseille. 
This work partly uses observations obtained with MegaPrime/MegaCam, a joint project of CFHT and CEA/DAPNIA, at the Canada-France-Hawaii Telescope (CFHT) which is operated by the National Research Council (NRC) of Canada, the Institut National des Sciences de l'Univers of the Centre National de la Recherche Scientifique (CNRS) of France, and the University of Hawaii. This work is based in part on data products produced at TERAPIX and the Canadian Astronomy Data Centre as part of the Canada--France--Hawaii Telescope Legacy Survey, a collaborative project of NRC and CNRS.
\end{acknowledgements}


\bibliographystyle{aa} 
\bibliography{paperbib} 


\end{document}